\documentclass[english,aps,twocolumn,prd,superscriptaddress,showpacs,amsmath,amssymbpreprintnumbers,]{revtex4-1}

\usepackage{dcolumn}
\usepackage{bm}
\usepackage{epsf}
\usepackage{amsfonts}
\usepackage{pifont} 
\usepackage[latin1]{inputenc}
\usepackage{babel}
\newcommand{\be}{\begin{equation}}
\newcommand{\ee}{\end{equation}} %\indent}
\newcommand{\eei}{\end{equation}\indent\indent}
\newcommand{\bc}{\begin{center}}
\newcommand{\ec}{\end{center}}
\newcommand{\ber}{\begin{eqnarray*}}
\newcommand{\ear}{\end{eqnarray*}}
\newcommand{\ba}{\begin{array}}
\newcommand{\ea}{\end{array}}
 
\newcommand{\na}{\nabla}

\newcommand{\hatn}{a}

\newcommand{\bea}{\begin{eqnarray}}
\newcommand{\eea}{\end{eqnarray}}
\newcommand{\nn}{\nonumber}

\newcommand{\ei}{\end{itemize}}

\newcommand{\D}{{\mathrm{D}}}
\newcommand{\e}{e}
\newcommand{\bra}[1]{\left(#1\right)}
\newcommand{\bras}[1]{\left[#1\right]}
\newcommand{\brac}[1]{\left\{#1\right\}}

\newcommand{\nab}{\nabla}
\newcommand{\la}{\langle}
\newcommand{\ra}{\rangle}

\newcommand \veps {\varepsilon} %curly epsilon

\newcommand{\MM}{{\cal M}}
\newcommand{\lb}{\{}%{\lceil}
\newcommand{\rb}{\}}%{\rfloor}
\newcommand{\A}{{\cal A}}

\newcommand{\E}{{\cal E}}
\renewcommand{\H}{{\cal H}}

\newcommand{\Lietwo}{{\cal L}}

\newcommand{\reff}[1]{(\ref{#1})}

\newcommand*\xbar[1]{%
  \hbox{%
    \vbox{%
      \hrule height 0.5pt % The actual bar
      \kern0.4ex%         % Distance between bar and symbol
      \hbox{%
        \kern-0.2em%      % Shortening on the left side
        \ensuremath{#1}%
        \kern-0.2em%      % Shortening on the right side
      }%
    }%
  }%
} 

\newcommand{\forget}[1]{\iffalse#1\fi}
    
\def\bal#1\eal{\begin{align}#1\end{align}}
\def\case#1/#2{\textstyle\frac{#1}{#2} }

\def\fp{f^{\prime}}
\def\fpp{f^{\prime \prime}}
\def\fppp{f^{\prime \prime \prime}}

\begin{document}
%%%%%%%%%%%%%%%%%%%%%%%%%%%%%%%%%%%%%%%%%%%%%
\title{Jebsen-Birkhoff theorem and its stability in $f(R)$ gravity}
%%%%%%%%%%%%%%%%%%%%%%%%%%%%%%%%%%%%%%%%%%%%%

\author{Anne Marie Nzioki}
\email{anne.nzioki@gmail.com}
\affiliation{Astrophysics Cosmology \&  Gravity Centre and 
Department of Mathematics and Applied Mathematics, 
University of Cape Town, Rondebosch,
7701, South Africa.}

\author{Rituparno Goswami}
\email{Goswami@ukzn.ac.za} 
\affiliation{Astrophysics \& Cosmology Research Unit, 
School of Mathematics Statistics and Computer Science,
University of KwaZulu-Natal, 
Private Bag X54001, Durban 4000, South Africa.}

\author{ Peter K. S. Dunsby}
\email{peter.dunsby@uct.ac.za}
\affiliation{Astrophysics Cosmology \&  Gravity Centre and 
Department of Mathematics and Applied Mathematics, 
University of Cape Town, Rondebosch,
7701, South Africa.}
\affiliation{South African Astronomical Observatory, 
Observatory, Cape Town, South Africa.}

%%%%%%%%%%%%%%%%%%%%%%%%%%%%%%%%%% 
\begin{abstract}
We prove a Jebsen-Birkhoff like theorem for $f(R)$ theories of gravity in order to to find the necessary conditions required for the existence of the Schwarzschild solution 
in these theories and demonstrate that the rigidity of such solutions of $f(R)$ gravity is valid even in the perturbed scenario.
\end{abstract}
%%%%%%%%%%%%%%%%%%%%%%%%%%%%%%%%%%
\pacs{}

\maketitle

%%%%%%%%%%%%%%%%%%%%%%%%%%%%%%%%%%
\section{Introduction}
%%%%%%%%%%%%%%%%%%%%%%%%%%%%%%%%%%

General Relativity (GR) has been one of the most successful theory in explaining the 
nature of gravity on both astrophysical and cosmological scales. However, with recent developments of high precision cosmology able to probe physics at a very large redshifts, the large scale validity of GR has come under increasing 
scrutiny. This is due to the fact that in order to fit the standard model of cosmology one has to introduce two dark components, namely the Dark Matter and the Dark Energy, in order to achieve a consistent picture. The problems related to these dark components remains to be one of the greatest puzzles in contemporary physics. One of the theoretical proposals that has received a considerable amount of attention is that Dark Energy has a geometrical origin. 
This idea is driven by the fact that modifications to GR appear in the low energy limit of many fundamental theories and that these 
modifications lead naturally to cosmologies which admit a Dark Energy like era without the introduction of any additional cosmological fields. 
The most popular candidates among these ultraviolet modifications of GR are the {\it fourth order} gravity theories, where the standard Einstein-Hilbert action 
of GR is modified by adding terms that lead to field equations of order four in the metric tensor. 

Although many of these modifications to GR has been somewhat successful in describing correctly the expansion history of the universe, on astrophysical scales there are considerable problems in modelling astrophysical objects like compact stars and black holes.
This is partially due to the added mathematical complexity of these theories and also due to the fact that in many of these theories the astrophysical objects become unstable, contrary to our own experience. Hence, in order to reach a viable alternative to General Relativity, one must do a detailed investigation on both astrophysical and cosmological scales.

We know in GR, spherically symmetric vacuum spacetimes 
have an extra symmetry: they are either locally static or spatially 
homogeneous. This rigidity of spherically symmetric vacuum solutions 
is the essence of {\it Jebsen-Birkhoff theorem} \cite{birkI,birkII,birkIII}. 
This theorem makes the Schwarzschild solution crucially important in 
astrophysics and underlies the way local astronomical systems decouple 
from the expansion of the universe \cite{CDGN}. The rigidity embodied in this 
property of the Einstein field equations is specific to vacuum GR solutions and is known not to hold for theories with extra degrees of freedom 
(e.g. {$f(R)$ theories of gravity or other scalar-tensor theories} \cite{scalar-tensor}). It is, therefore, important to investigate the extra conditions required for a Jebsen-Birkhoff like theorem to hold modified theories of gravity.

It was recently shown \cite{amostBirk,amostBirkmatter,varBirk}, that in GR, the rigidity of spherical vacuum solutions of Einstein's field equations continues 
even in the perturbed scenario: almost spherical symmetry and/or almost 
vacuum imply almost static or almost spatially homogeneous. This provides an 
important reason for the stability of the solar system and of black hole spacetimes 
and has interesting implications for the issue of how a universe made up of locally spherically symmetric objects imbedded in vacuum regions is able 
to expand, given that Birkhoff's theorem tells us the local spacetime domains 
have to be static. A similar study of local stability is required for the spherically 
symmetric solutions in modified gravity theories to see if these theories are 
physically viable. 

In this paper, we prove a Jebsen-Birkhoff like theorem for $f(R)$ theories of gravity, 
to find the necessary conditions required for the existence of a Schwarzschild solution 
in these theories. We discuss under what circumstances we can covariantly set up a 
scale in the problem and then perturb the vacuum spacetime with respect to this 
covariant scale to find the stability of the theorem. We do this in two steps: 
(a) First we maintain spherical symmetry and perturb the Ricci scalar around $R=0$ 
to find the necessary conditions on the spatial and temporal derivatives of the 
Ricci scalar for the spacetime to be almost Schwarzschild.
(b) We then define the notion of {\em almost spherical symmetry} with respect to 
the covariant scale and perturb the spherical symmetry to prove the stability of the 
theorem.

The important result that emerges covariantly from this investigation is that, there 
exists a non-zero measure in the parameter space of $f(R)$ theories for which 
the Jebsen-Birkhoff like theorem remains stable under generic perturbations.Furthermore our result applies locally and hence does not depend on specific 
boundary conditions used for solving the perturbation equations. We prove the 
result by using the 1+1+2 covariant perturbation formalism \cite{extension,Gary,chris,BCMD}, 
which developed from the 1+3 covariant perturbation formalism \cite{Covariant}.

%%%%%%%%%%%%%%%%%%%%%%%%%%%%%%%%%%%%%%%%%
\section{General equations for fourth order gravity}
%%%%%%%%%%%%%%%%%%%%%%%%%%%%%%%%%%%%%%%%% 
The simplest generalisation of the Einstein-Hilbert action of GR is obtained by 
replacing the Ricci scalar $R$ by a function of the Ricci scalar $f(R)$, resulting 
in the action
\be
{\cal A}= \frac12 \int d^4x  \bras{\sqrt{-g}\,f(R)+ \Lietwo_{m}}\;,
\label{FOGaction}
\ee
where $ \Lietwo_{m}$ is the Lagrangian density of the standard matter fields. 
Varying the action with respect to the metric over a 4-volume gives the following 
field equations
\be
G_{ab} = \bra{R_{ab}-\frac12g_{ab}R} = T_{ab}
= \frac{T^{M}_{ab}}{\fp} + T^{R}_{ab}\;,
\label{fieldeq} 
\ee
where the right hand side is the ``effective'' energy momentum tensor $T_{ab}$ 
comprising $ T^{M}_{ab} $, the standard matter energy momentum tensor and \be
T^{R}_{ab} = \frac{1}{\fp} \bras{\frac12 g_{ab} \bra{f-R\, \fp}
+ \nabla_{b}\nabla_{a} \fp- g_{ab}\nabla_{c}\nabla^{c} \fp} ~,
\ee 
which we label the ``curvature fluid'' energy momentum tensor.  

%%%%%%%%%%%%%%%%%%%%%%%%%%%%%%%%%%%%%%%%% 
\section{1+1+2 Covariant splitting of spacetime}
%%%%%%%%%%%%%%%%%%%%%%%%%%%%%%%%%%%%%%%%% 
%%%%%%%%%%%%%%%%%%%%%%%%%%%%%%%%%%%%%%%%% 
\subsection{Kinematics}
%%%%%%%%%%%%%%%%%%%%%%%%%%%%%%%%%%%%%%%%% 
In the 1+3 covariant formalism {\cite{Covariant}} a non-intersecting timelike family 
of worldlines (associated with \textit{fundamental observers} comoving with the 
cosmological fluid) forms a congruence in spacetime ($\MM$, \textbf{g}) 
representing the average motion of matter at each point. The congruence is defined by a timelike unit vector $u^a$ ($u^a u_a =-1$) splitting the spacetime 
in the form $R\otimes V$ where $R$ denotes the timeline along $u^a$ and $V$ is the tangent 3-space perpendicular to $u^a$. The projection tensor 
\be
h^a{}_b=g^a{}_b+u^a \,u_b~, \quad h^{a}{}_{a} = 3~,
\label{proj1+3} 
\ee 
projects into the rest space orthogonal to $u^a$ and the projected alternating 
Levi-Civita tensor $\veps_{abc}$ is the effective volume element for the 3-space.  

Any spacetime 4-vector $\psi_{a} $ may be covariantly split into a scalar, $\psi$, 
which is the part of the vector parallel to $u_{a} $, and a 3-vector, $\psi_{a}$, 
lying orthogonal to $u_{a} $:
\be
\psi_{a} = - \psi \,u_{a} + \psi_{\la a \ra}~,
 \quad  \psi \equiv \psi_{b}\,  u^{b}~, 
 \quad \psi^{\la a \ra} \equiv h^{a}{}_{b}\, \psi^{b}~.  
  \label{equation1}
\ee
and any projected rank-2 tensor $ \psi_{cd}$ can be split as 
\be
\psi_{ab} = \psi_{\la ab \ra}  + \frac13 \psi\,  h_{ab}  + \psi_{[ab]}~, \label{equation2}
\ee
where $ \psi = {h}_{cd}  \psi^{cd} $ is the spatial trace, 
$\psi_{\la ab \ra}$ is the orthogonally \textit{projected symmetric trace-free} 
PSTF part of the tensor defined as
\be
\psi_{\la ab \ra} = \bra{h_{(a}{}^{c} \,  h_{b)}{}^{d} 
- \frac{1}{3} h_{ab}\, h^{cd}} \psi_{cd} ~,
\label{PSTF1+3}
\ee
and $\psi_{[ab]}$ is the skew part of the rank-2 tensor which is spatially dual to the spatial vector $\psi^{c} $ ($\psi_{[ab]} = \veps_{abc} \, \psi^{c }$). 
The angle brackets denote orthogonal projections of vectors and the 
orthogonally PSTF part of tensors.

Moreover, two derivatives can be defined: the vector $ u^{a} $ is used to define the \textit{covariant time derivative} 
(denoted with a dot - $`\dot{\phantom{x}}$') along the observers' worldlines, 
where for any tensor $ S^{a..b}{}_{c..d} $    
\be
\dot{S}^{a..b}{}_{c..d}= u^{e}\,  \nab_{e} {S}^{a..b}{}_{c..d}~, 
\ee
and the spatial projection tensor $ h_{ab} $ is used to define the fully 
orthogonally projected \textit{covariant spatial derivative} `$\D$', such that, 
\be
\D_{e}S^{a..b}{}_{c..d}= h^r{}_e\, h^a{}_f  \,...\,  
h^b{}_g\, h^p{}_c \,...\, h^q{}_d \,\na_{r} {S}^{f..g}{}_{p..q}~, 
\ee
with the projection on all the free indices. 

In the 1+1+2 approach, we further split the 3-space $V$, by introducing the spacelike 
unit vector $ e^{a} $ orthogonal to $u^{a} $ so that 
\be 
e_{a} u^{a} = 0\;,\; \quad e_{a} e^{a} = 1~. 
\ee
Then the \textit{projection tensor} 
\be 
N_{a}{}^{b} \equiv
h_{a}{}^{b} - e_{a}e^{b} ~, \quad N^{a}{}_{a} = 2~, 
\label{proj1+1+2} 
\ee 
projects vectors onto the tangent 2-surfaces orthogonal to 
$e^{a}$ \textit{and} $u^a$, which, following \cite{extension}, 
we will refer to as `{\it sheets}'. The sheet carries a natural 2-volume element:
\be
\veps_{ab}\equiv\veps_{abc}e^{c}~.
\label{perm1+2}
\ee

In 1+1+2 slicing, any 3-vector $ \psi_{\la a \ra}$ as defined in 
\reff{equation1}, can be irreducibly split into a component along $e^{a}$ and 
a sheet component $\Psi^{a}$, orthogonal to $e^{a}$ i.e. 
\be 
\psi^{\la a \ra} = \Psi \,e^{a} + \Psi^{a}\,,
\quad \Psi\equiv\psi_{\la a \ra}\, e^{a}\,, \quad\Psi^{a} \equiv
N^{ab}\psi_{\la b \ra}~. 
\label{equation3} 
\ee
and a similar decomposition can be done for a PSTF 3-tensor 
$\psi_{\la ab \ra}$ as defined \reff{PSTF1+3}, which can be split into
scalar along $e^a$, a 2-vector and a 2-tensor part as follows: 
\be
\psi_{ab} = \psi_{\la ab\ra} = \Psi\bra{e_{a}e_{b} -
\frac{1}{2}N_{ab}}+ 2 \Psi_{(a}e_{b)} + \Psi_{ab}~,
\label{equation4} 
\ee 
where 
\bea
\Psi &\equiv & e^{a}e^{b}\psi_{ab} = -N^{ab}\psi_{ab}~,\nn \\
\Psi_{a} &\equiv & N_{a}{}^be^c\psi_{bc}~,\nn \\
\Psi_{ab} &\equiv & \psi_{\brac{{ab}}}
= \bra{ N^{c}{}_{(a}N_{b)}{}^{d} - \frac{1}{2}N_{ab} N^{cd}} \psi_{cd}~,
\eea
and the curly brackets denote the PSTF part of a tensor with respect to $e^{a}$. 

Apart from the `{\it time}' (dot) derivative of an object (scalar, vector or tensor), 
in the 1+1+2 formalism we introduce two new derivatives, which $ e^{a} $ 
defines, for any object $ \psi^{a...b}{}_{c...d}  $:
\begin{align}
\hat{\psi}^{a...b}{}_{c...d}  &\equiv 
e^{f}D_{f}\psi^{a...b}{}_{c...d} ~,
\label{hatdrv}\\
\delta_f\,\psi^{a...b}{}_{c...d} &\equiv 
N^j{}_f \,N^{a}{}_{i} \,...\,N^{b}{}_{j}  \, N^{l}{}_{c} \,...\, 
N^{g}{}_{d}  \,D_j\psi^{i..j}{}_{l..g}\;. 
\label{deltadrv}
\end{align}
The hat-derivative $`\hat{\phantom{x}}$' is the derivative along the $e^a$ 
vector-field in the surfaces orthogonal to $ u^{a} $. The $\delta$-derivative is 
the projected derivative onto the orthogonal 2-sheet, with the projection on 
every free index. 

The fundamental geometrical quantities in the spacetime in the 1+1+2 
formalism for $f(R)$ gravity are (see \cite{chris} for a detailed physical description of these variables),
\bea
\left[ R, \Theta, \A, \Omega,\Sigma, \E, \H, \phi,\xi,
\A^{a},\Omega^{a}, \Sigma^{a},   \right. \nonumber\\
\left. \alpha^a, a^a,\E^{a}, \H^{a}, \Sigma_{ab}, \E_{ab}, \H_{ab},\zeta_{ab} \right] \,,
 \label{geosplit}
\eea
and their dynamics give us information about the spacetime geometry.

In terms of these variables, the expression for the full covariant derivative of 
$\e^{a}$ in its irreducible form is
\bea
\na_a\,\e_b&=&-\A\, u_a\,u_b-u_a\,\alpha_b + \bra{\frac13\Theta
+\Sigma}\e_{a}\,u_{b}  +\xi\,\veps_{ab}+\zeta_{ab}\nn\\
&& + \bra{\Sigma_a-\veps_{ac}\,\Omega^c}\,u_b+\e_a\,\hatn_b 
+\frac12\phi\,N_{ab}\ , 
\label{fullcov_e}
\eea
from which we can obtain the spatial derivative of $\e^{a}$ as
\be 
{\rm \D}_{a}\e_{b} = \e_{a}\, \hatn_{b} + \frac{1}{2}\phi \,N_{ab} 
+ \xi \,\veps_{ab} + \zeta_{ab}~.
 \label{cov_der}
\ee
The other derivative of $ \e^{a} $ is its change along $ u^{a} $,
\be 
\dot{\e}_{a} = \A \,u_{a} + \alpha_{a}
~~\mathrm{where}~~ \A \equiv \e^a{\dot{u}_{a}} 
~~ \mathrm{and}~~\alpha_{a}\equiv N_{a}{}^{b}\dot{\e}_{b}~.
\ee 
From equation \reff{cov_der} we see that along the spatial direction 
$e^{a}$, $ \phi =  \delta_{a}\e^{a}$ represents the expansion of the sheet, 
$\zeta_{ab} = \delta_{\lb a}\e_{b \rb }$ is the shear of $e^{a}$ (i.e., the distortion 
of the sheet) and $ \hatn_a = e^c\,\D_c\, \e_a= \hat \e_a$ its acceleration, while 
$\xi = \frac{1}{2}\, \epsilon^{ab}\delta_{a}\e_{b}$ represents the 
vorticity associated with $e^{a}$ (`twisting' of the sheet). 

We include here the expression for the 1+1+2 split of the full covariant 
derivative of $u^{a}$ 
\bea
\na_a u_b&=&
-u_a\bra{\A\,\e_b+\A_b}+\e_a\,\e_b\bra{\frac13\theta+\Sigma}\nn\\&&
+\e_a\bra{\Sigma_b+\veps_{bc}\Omega^c}
+\bra{\Sigma_a-\veps_{ac}\Omega^c}\e_b\nn\\&&
+N_{ab}\bra{\frac13\theta-\frac12\Sigma}+
\Omega\,\veps_{ab}+\Sigma_{ab}~, 
\label{fullcov_u}
\eea

In general the three derivatives defined so far, dot - $`\dot{\phantom{x}}$', 
hat - $`\hat{\phantom{x}}$' and delta - $`\delta_a$', do not commute. 
The commutations relations for these derivatives of any scalar $\psi$ are
\begin{align}
\hat{\dot \psi}-\dot{\hat \psi} =&-\A\dot\psi+\bra{\frac13 \Theta 
+ \Sigma}\hat\psi+ \bra{\Sigma_a + \veps_{ab}\Omega^b
-\alpha_a}\delta^a\psi ~,
\label{comm1} 
\end{align}
\begin{align}
\delta_a\dot\psi-\bra{\delta_a\psi}^{~\cdot}_{\perp} =&
-\A_{a}\,\dot\psi + \bra{\alpha_{a} +\Sigma_a 
- \veps_{ab}\Omega^b}\hat\psi \nn \\
&+ \bra{\frac13\Theta-\frac12\Sigma}\delta_a\psi 
+ \bra{\Sigma_{ab}+\Omega\,\veps_{ab}}\delta^b\psi\  ~, 
\label{comm2} 
\end{align}
\begin{align}
\delta_a\hat\psi-{\bra{\delta_a\psi}}^{~\hat{}}_{\perp}  =& 
-2\,\veps_{ab}\Omega^b\,\dot\psi+\hatn_a\,\hat\psi
+\frac12\phi\,\delta_a\psi \nn \\
&+\bra{\zeta_{ab}+\xi\,\veps_{ab}}\delta^b\psi\  ~,
 \label{comm3} 
 \end{align}
\begin{align}
\delta_{[a}\delta_{b]}\psi =&\veps_{ab}\bra{\Omega\, \dot\psi-\xi\,\hat\psi} 
+  \hatn_{[a}\delta_{b]} \psi~.\label{comm4}
\end{align}
From the above relations it is clear that the 2-sheet is a genuine 2-surface 
(instead of just a collection of tangent planes) if and only if the commutator 
of the time and hat derivative do not depend on any sheet component, 
i.e., when $\Sigma_a + \veps_{ab}\Omega^b-\alpha_a =0$ and the sheet 
 derivatives commute, i.e., when $\xi=\Omega=\hatn_a=0$.

%%%%%%%%%%%%%%%%%%%%%%%%%%%%%%%%%%%%%%%%% 
\subsection{Energy momentum tensor}
%%%%%%%%%%%%%%%%%%%%%%%%%%%%%%%%%%%%%%%%% 
In terms of the 1+1+2 variables, fluid description of the energy momentum tensor 
in \reff{fieldeq} is given by
\bea
T_{ab} &=& \mu\, u_{a} u_{b} + p \bras{N_{ab} + e^{a}e^{b}}
		+ 2 u_{(a}\bras{Q \,e_{b)} +Q_{a)}} \nonumber\\  
	&&+ \,\Pi \bras{e_{a}\,e_{b}-\frac12 N_{ab}}+2 \Pi_{(a} e_{b)}+\Pi_{ab} ~.
 \label{EMTsplit}
\eea
We recall here that the ``effective'' thermodynamic quantities as presented in 
\reff{EMTsplit} are representative of the total combination of the standard matter and 
curvature quantities as follows:
\begin{align}
\mu =&\, \mu^{M} + \frac{1}{\fp} \bras{\frac12 (R \fp-f) - \theta  \fpp \dot{R} 
+ \fppp  X^{2}+  \fpp  \hat{ X}  \right. \nn \\
&\left.+ \,  \fppp \delta^{a} R \delta_{a} R  + \phi \fpp \, X  
 - \hatn^{a} \fpp  \delta_{a}R + \fpp \delta^{a}\delta_{a} R  } ~, 
 \label{generaldensity}
 \end{align}
 \begin{align}
p =&\, p^{M} +  \frac{1}{\fp} \bras{\frac12 (f-R\, \fp) + \frac23 \theta  \fpp \dot{R}
+ \fppp \dot{R}^{2} + \fpp \ddot{R}   \right. \nn \\  
&	\left.-\,   \A \fpp\, X- \A^{a} \fpp  \delta_{a} R  -\frac23( \phi  \fpp X 
+ \fppp  \delta^{a} R \delta_{a} R  \right. \nn \\  
& \left. +\, \fpp \delta^{a} \delta_{a} R + \fppp  X^{2} +  \fpp\hat{X} 
- \hatn_{a}  \fpp  \delta^{a}R)} ~,
 \label{generalpressure}
  \end{align}
 \begin{align}
Q=& \,Q^{M} - \frac{1}{\fp} \bras{  \fppp  \dot{R} X 
+ \fpp \bra{ \dot{X}-\A \dot{R} } -\alpha^{a} \fpp\delta_{a}R } ~,
 \label{generalfluxSc}
  \end{align}
 \begin{align}
Q_{a}=& \, Q_{a}^{M} + \frac{1}{\fp} \bras{\bra{ \Sigma_{a} 
-\veps_{a}{}^{b}  \Omega_{b}} \fpp  X 
- \fpp \delta_{a} \dot{R} \right. \nn \\  
& \left. +\,  \bra{ \Sigma_{a}{}^{b} + \veps_{a}{}^{b} \Omega} \fpp  \delta_{b} R   
-\dot{R} \fppp \delta_{a}R  \right. \nn \\  
& \left.+\, \bra{ \frac13 \theta
-\frac12 \Sigma}  \fpp  \delta_{a} R }  ~,  
 \label{generalfluxVc}
  \end{align}
 \begin{align}
\Pi =& \, \Pi^{M}  + \frac{1}{\fp} \bras{ \frac13 \bra{ 2 \fppp X^{2} 
+2 \fpp \hat{X} - 2 \A_{a}\, \fpp  \delta^{a} R  
\right. \right. \nn \\  
& \left. \left.- \, \phi \fpp  X- \fppp  \delta^{a}R \delta_{a} R  
- \fpp  \delta^{a} \delta_{a} R } -  \Sigma   \fpp \dot{R} } ~,
 \label{generalstressSc}
  \end{align}
 \begin{align}
\Pi_{a} =&\, \Pi_{a}^{M} +  \frac{1}{\fp} \bras{ - \Sigma_{a}\, \fpp  \dot{R} 
+X \fppp  \delta_{a} R + \fpp  \delta_{a} X \right. \nn \\  
&\left. -\,  \frac12 \phi  \fpp  \delta_{a} R +\bra{ \xi \veps_{a}{}^{b}
- \zeta_{a}{}^{b} }\fpp \delta_{b} R 
 \right. \nn \\  
&\left.-\,  \frac12 \bra{ \Sigma_{a} 
+ \veps_{a}{}^{b} \Omega_{b}} \fpp \dot{R}   } ~,
 \label{generalstressVc}
  \end{align}
 \begin{align}
\Pi_{ab} =& \,\Pi_{ab}^{M} +\frac{1}{\fp} \bra{- \Sigma_{ab}  \fpp \dot{R} 
+ \zeta_{ab}  \fpp X \right. \nn \\  
&\left. +  \, \fppp \delta_{\lb a} R \delta_{b \rb} R  
+  \fpp  \delta_{\lb a} \delta_{b \rb} R} ~,
 \label{generalstressTn}
\end{align}
where $\mu^M$ is the energy density relative to $ u^{a}$, $p$ the isotropic pressure, 
$Q$ and $Q_{a}$ are the components of the $ u^{a}$ energy flux parallel 
and orthogonal to $e^{a}$ respectively, $\Pi$, $\Pi_{a}$ and $\Pi_{ab}$ are the PSTF 
( w.r.t $ e^{a}$) parts of the anisotropic pressure. We define also define $\hat{R}=X$.

The set of thermodynamic matter variables,
\be
\{\mu^{M},\, p^{M},\, Q^{M},\,\Pi^{M},\,Q^{M}_a,\,\Pi^{M}_a,\,  \Pi^{M}_{ab} \}~, 
\label{thersplit}
\ee
for a given equation of state, together with \reff{geosplit} form an {\it irreducible set} 
that completely describes the vacuum spacetime in $f(R)$ gravity. For the complete 
set of evolution equations, propagation equations, 
mixed equations and constraints for the above irreducible set of variables please see equations (48-81) of \cite{chris}.

%%%%%%%%%%%%%%%%%%%%%%%%%%%%%%%%%%%
\section{1+1+2 equations for vacuum LRS-II spacetimes} 
%%%%%%%%%%%%%%%%%%%%%%%%%%%%%%%%%%%
A spacetime is said to be locally rotationally symmetric (LRS) if there exists a 
continuous isotropy group at each point and hence is characterised by the existence 
of a multi-transitive isometry group acting on the spacetime manifold 
\cite{EllisLRS, StewartLRS, MarklundLRS}. These spacetimes exhibit locally 
(at each point) a unique preferred spatial direction that constitutes a local axis 
symmetry such that the geometry is invariant under rotations about it. We choose 
the vector field $ e^{a} $ as the preferred spatial direction in the LRS spacetime, 
namely the `radial' vector. Now since LRS spacetimes are defined to be isotropic, 
this allows for the vanishing of \textit{all} 1+1+2 vectors and tensors, such that there 
are no preferred directions in the sheet. Thus, in vacuum 
($\mu^M= p^M=Q^M=\Pi^M=0$), all the non-zero 1+1+2 variables are the covariantly 
defined scalars
\be
\textbf{LRS}: ~ \brac{R,\, \A, \,\Theta,\,\phi, \,\xi, \,\Sigma,
\,\Omega,\, \E, \,\H}\ ,
\label{LRSvars}
\ee
A detailed discussion of the covariant approach to LRS perfect fluid space-times can 
be seen in \cite{EllisLRS}.

One subsets of LRS spacetimes is the LRS class II, which contains all the LRS spacetimes 
that have no vorticity or spatial rotation. As a consequence, the vorticity 
components $\Omega$ and  $\xi$ associated with $u^a$ and $e^a$, respectively, and 
$ \H $ which is a component of the magnetic Weyl curvature (all these quantities are in 
the surfaces orthogonal to $u^a$), are identically zero in the LRS-II spacetimes. The 
set of remaining variables are
\be
\textbf{LRS class II}: ~ \brac{R, \A, \,\Theta,\,\phi,
 \,\Sigma,\,\E }\ ,
\ee
where $\Theta$ the 3-volume rate of expansion, $\Sigma$ is the component of shear 
parallel to $e^{a}$ and $\E$ is the component of the electric Weyl tensor, also parallel 
to $e^{a}$. These quantities  fully characterise the kinematics and dynamics of the 
LRS II spacetime and their dynamics, based on the Ricci and Bianchi identities, is 
governed by the following equations :

\subsection*{Propagation equations}
\bea 
\hat\phi&= -&\frac12\phi^2+\bra{\frac13\Theta
+\Sigma}\bra{\frac23\Theta-\Sigma}\nonumber\\
&&-\frac23\mu -\frac12\Pi-\E\;, \label{hatphi}\\
\hat\Sigma-\frac23\hat\Theta&= -&\frac32\phi\Sigma-Q~,
\\
\hat\E-\frac13\hat\mu+\frac12\hat\Pi&= -&\frac32\phi\bra{\E+
\frac12\Pi}+ \bra{\frac12\Sigma-\frac13\Theta}Q~. \label{Ehat}
\nonumber \\
\eea
\subsection*{Evolution equations}
\bea
\dot\phi &= -&
\bra{\Sigma-\frac23\Theta}\bra{\A-\frac12\phi} +Q~,\label{Q}
\\
\dot\Sigma-\frac23\dot\Theta&= -&\A\phi +2\bra{\frac13\Theta
-\frac12\Sigma}^{2}\nonumber\\
&&+\frac13 \bra{\mu+3p}-\E +\frac12\Pi~,
\\ 
\dot\E-\frac13\dot\mu+\frac12\dot\Pi&= &\bra{\frac32\Sigma-\Theta}\E
+\frac14\bra{\Sigma-\frac23\Theta}\Pi\nonumber\\&&
+\frac12\phi Q-\frac12\bra{\mu+p}\bra{\Sigma-\frac23\Theta}. 
\label{Edot} 
\eea 
\subsection*{Propagation/Evolution Equations}
\bea
\dot\mu+\hat Q&= -&\Theta\bra{\mu+p}-\bra{\phi+2\A}Q -
  \frac32\Sigma\Pi~,
\\ 
\dot Q +\hat p+\hat\Pi&= -&\bra{\frac32\phi+\A}\Pi-
\bra{\frac43\Theta+\Sigma} Q \nonumber\\
&&-\bra{\mu+p}\A~,
\\
\hat\A-\dot\Theta&= -&\bra{\A+\phi}\A+\frac13\Theta^2 \nonumber\\
&&   +\frac32\Sigma^2 + \frac12\bra{\mu+3p}~. 
\label{ray1}
\eea
where
\ber
\mu	&=& \frac{1}{f'} \bras{\frac12 (Rf'-f)    -  \theta  f''\dot{R}  +  f''' X^{2}  
			+  f'' \hat{ X} + \phi f'' X  } ~, \nonumber\\
p &=& \frac{1}{f'} \bras{\frac12 (f-Rf') 
			+ f'''\dot{R}^{2} + f''\ddot{R}  - \A f'' X \right. \nonumber\\  
		&&\left. + \frac23 \bra{ \theta  f'' \dot{R} - \phi  f'' X   -  f''' X^{2} 
		-  f'' \hat{X} }}  ~,\nonumber \\
Q &=& - \frac{1}{f'} \bras{  f''' \dot{R} X + f''\bra{ \dot{X}-\A \dot{R} } }  ~,
			\nonumber \\
\Pi &=& \frac{1}{f'} \bras{ \frac13 \bra{ 2 f''' X^{2} +2 f'' \hat{X} 
			- \phi f'' X  } -  \Sigma  f'' \dot{R} } ~.
\ear
\subsection*{Commutation relation}
\be
\hat{\dot{\psi}} - \dot{\hat{\psi}} = -\A \dot\psi + 
\bra{\frac13\Theta+\Sigma}\hat{\psi}\;. \label{commutation}
\ee 
Due to the additional degrees of freedom, equations 
\reff{hatphi}-\reff{commutation} are not closed and we have to add 
an additional equation that we label the {\it trace equation}:
\bea
R f'  - 2 f  &=&  3\bra{  f'' \theta  \dot{R} -f''\hat{ X}  +  f''\ddot{R} -   \bra{\phi+\A} f'' X\right. \nonumber\\  
		&&\left. 
		-  f''' X^{2}+  f'''\dot{R}^{2} }
			\label{trace_SS} 
\eea
Since the vorticity vanishes, the Gauss equation for $ e^{a} $ together 
with the 3-Ricci identities determine the 3-Ricci curvature tensor of the 
spacelike 3-surfaces orthogonal to $ u^{a} $ to be  \cite{Gary}:
\be
^{3}R_{ab} = -\bras{\hat{\phi}+\frac12 \phi^{2}}e_{a}\,e_{b} - 
\bras{\frac12 \hat{\phi} + \frac12\phi^{2} - K}N_{ab}~.
\ee
This gives the 3-Ricci-scalar as
\be
^{3}R = -\,2\bras{\frac12 \hat{\phi} + \frac34\phi^{2} - K}\ ,
\label{3nRicci}
\ee
where $ K $ is the \textit{Gaussian curvature} of the 2-sheet and is related to 
the two dimensional Riemann curvature tensor and two dimensional Ricci 
tensor as
\be
{^{(2)}}R^a{}_{bcd} = K \bra{N^a{}_c \,N_{bd} - N^a{}_d\,N_{bc}}~,
~ \Longrightarrow ~ ^{2}R_{ab}=K\,N_{ab}~.
\label{2-RiemRic} 
\ee 
From \reff{3nRicci} and \reff{hatphi} an expression for $ K $ is obtained in the form
\be
K = \frac13 \mu - \E - \frac12 \Pi + \frac14 \phi^{2} - 
\bra{\frac13 \Theta - \frac12 \Sigma}^{2} ~.
\label{GaussCurv}
\ee
From \reff{hatphi}-\reff{Edot}, we obtain the evolution and propagation equations 
of $K$ as 
\bea
\dot{K} = - \bra{\frac23 \Theta - \Sigma}K~,
\label{evoGauss}\\ 
\hat{K} = -\phi K~.
\label{propGauss}
\eea
From equation \reff{evoGauss}, it follows that whenever the Gaussian curvature of the 
sheet is non-zero and constant in time, then the shear is always proportional to the 
expansion as 
\be 
K \neq 0\quad \textrm{and}\quad \dot K =
0\quad \Longrightarrow \quad \Sigma = \frac23\Theta\ ~. 
\label{prop}
\ee

%%%%%%%%%%%%%%%%%%%%%%%%%%%%%%%%%%%
\section{Symmetries} 
%%%%%%%%%%%%%%%%%%%%%%%%%%%%%%%%%%%
We know geometrically LRS-II space times have some inherent symmetries that lie 
on the 2-sheets. To investigate the extra symmetry of vacuum LRS-II space times 
for the modified theories, we follow \cite{amostBirk}, by trying to solve the {\it Killing equation} for a Killing vector of the 
form $\xi_a=\Psi\, u_a+\Phi \,e_a$, where $\Psi$ and $\Phi$ are scalars. The Killing 
equation gives 
\be 
\na_a(\Psi\,u_b+\Phi \,e_b) + \na_b(\Psi\, u_a+\Phi \,e_a) =0\;. 
\label{Killing} 
\ee
Using equations \reff{fullcov_e} and \reff{fullcov_u}, and multiplying the Killing 
equation by $u^a\,u^b$, $u^a\,e^b$, $e^a\,e^b$ and $N^{ab}$ results in the following 
differential equations and constraints: 
\bea
\dot\Psi+\A\,\Phi&=&0\;, 
\label{psidot}\\
\hat\Psi -\dot\Phi-\Psi\,\A+\Phi\bra{\Sigma+\frac13\Theta}&=& 0\;,
\label{psihat}\\
\hat\Phi+\Psi \bra{\frac13\Theta+\Sigma}&=&0\;,
\label{cons1}\\
\Psi\bra{\frac23\Theta-\Sigma}+\Phi\,\phi&=&0\;.
\label{cons2}
\eea
Considering that $\xi_a\,\xi^a=-\Psi^2+\Phi^2$, if $\xi^a$ is timelike 
(that is, $\xi_a\,\xi^a<0$), then because of the arbitrariness in choosing the vector 
field $u^a$, we can always make $\Phi=0$, while if $\xi^a$ is spacelike (that is  $\xi_a\,\xi^a>0$), then we can make $\Psi=0$.

Let us first assume that $\xi^a$ is timelike and $\Phi=0$. Looking at \reff{psidot} 
and \reff{psihat}, we know that their solutions always exists. For a non trivial $\Psi$, 
the constraints \reff{cons1} and \reff{cons2} together imply, that in general 
$\Theta=\Sigma=0$, that is, the expansion and shear of a unit vector field along the 
timelike Killing vector vanishes. We also see that the time derivatives of all the 
quantities in the field equations \reff{hatphi}-\reff{trace_SS} vanish and hence the 
spacetime is \textit{static}. 

Now if $\xi^a$ is spacelike and $\Psi=0$, then we see in 
this case that solution of equations \reff{psihat} and \reff{cons1} always exists and 
the constraints \reff{psidot} and \reff{cons2} together imply that in general, (for a 
non trivial $\Phi$), $\phi=\A=0$. If we impose further the condition,
\be
R = R_{0} = const. \quad \mathrm{and} \quad \fp_{0} \neq 0 ~,
\label{cond}
\ee
which in turn implies
\bea
\Pi	&=& 0 ~,
\\
\mu	&=& \frac{1}{\fp_{0}} \bras{\frac12 (R_{0}\,\fp_{0}-f_{0}) } ~,
\\ 
p 	&=& \frac{1}{\fp_{0}} \bras{\frac12 (f_{0} - R_{0}\, \fp_{0})  }  ~, 
\\
R_{0}\, \fp_{0}  - 2 f_{0} &=&0~,
\eea
where $\fp(R_0)=\fp_0$, then all the spatial derivatives of all the quantities in 
\reff{hatphi}-\reff{trace_SS} vanish. From this we see that \textit{homogeneity} 
is only achieved if $R =  constant $, otherwise inhomogeneity is admitted for 
non-constant $R$. This result is unlike that of GR where the spacetime is spatially 
homogenous upon setting $\phi=\A=0$ in the list of LRS II equations. 

We can now say that : \textit{There always exists a Killing vector in the local $[u,e]$ 
plane for a vacuum LRS-II spacetime in $f(R)$ gravity. If the Killing vector is timelike 
then the spacetime is locally static. If the Killing vector is spacelike, the spacetime is 
{\it locally spatially homogeneous} if and only if 
$R = R_{0} =  const. $ and $ \fp_{0} \neq 0$.}

%%%%%%%%%%%%%%%%%%%%%%%%%%%%%%%%%%%
\section{Jebsen-Birkoff like theorem in $f(R)$ gravity} 
%%%%%%%%%%%%%%%%%%%%%%%%%%%%%%%%%%%
If we apply the conditions \reff{cond}, to the system of  equations 
\reff{hatphi}-\reff{trace_SS}, then from \reff{Ehat}, \reff{Edot}, \reff{evoGauss} 
and \reff{propGauss} we get
\be 
\E=C\,K^{3/2}\ . 
\label{EK} 
\ee
That is, the 1+1+2 scalar of the electric part of the Weyl tensor is always proportional 
to the (3/2)th power of the Gaussian curvature of the 2-sheet. The proportionality 
constant $C$ sets up a scale in the problem in this particular case.

Furthermore, when $\Theta=\Sigma=0$, we have $\dot K=0$. We choose 
coordinates to make the Gaussian curvature `$K$' of the spherical sheets proportional 
to the inverse {square} of the radius co-ordinate `$r$', (such that this coordinate becomes 
the {\it area radius} of the sheets), then this geometrically relates the `{\it hat}'  derivative 
with the radial co-ordinate `$r$'. Using \reff{propGauss}, \reff{hatdrv} and the definition of 
$ \phi$ we can then define the hat derivative of any scalar $M$ as 
\be 
\hat{M}=\frac{1}{2}r\phi\frac{d M}{d r}
\label{hatr} 
\ee   
for a static spacetime. 

If we take $R =R_{0} =  0, f(0) = 0$ and $ f'_{0} \neq 0$, the equations 
\reff{hatphi}-\reff{trace_SS} reduce to 
\bea
\hat\phi &=& - \frac12 \phi^{2} - \E 
\label{schwphi} \\
\hat \E &=& - \frac32  \phi \E
 \label{schwE} 
\eea
together with the constraint:
\bea
\E + \A \phi = 0 ~. 
\label{schwconstraint1}
\eea
The local Gaussian curvature of the 2-sheets in this case becomes,
\be
K = - \E + \frac14 \phi^{2}~.
\label{GaussCurv1}
\ee
The parametric solutions for these variables (when $K>0$ ) are
\bea
\phi=\frac{2}{r}\sqrt{1-\frac{2m}{r}}\;&,&\;\A=\frac{m}{r^2}~,
\left[1-\frac{2m}{r}\right]^{-\frac12}\\
\E=\frac{2m}{r^3}\;&,&\;K=\frac{1}{r^2}~,
\label{Schw1}
\eea
where $m$ is the constant of integration. 

Solving for the metric using the definition of these geometrical quantities we get ~\cite{Gary}
\be
ds^2=-\left(1-\frac{2m}{r}\right)dt^2+\frac{dr^2}{(1-\frac{2m}{r})}
	+r^2d\Omega^2,
\ee
which is the metric of a static Schwarzschild exterior. A similar derivation can 
be done for the case of spacelike Killing vector and vanishing Ricci scalar, 
to get the Schwarzschild interior metric. 

We can now give a generalisation of the {Jebsen-Birkhoff-like theorem in $f(R)$} gravity:

\textit{For $f(R)$ gravity, where the function $f$ is of class $C^3$ at $R=0$,with $f(0) = 0$ 
and $\fp(0)\ne 0$, the only spherically symmetric solution with vanishing Ricci scalar in 
empty space in an open set $\mathcal{S}$, is one that is locally equivalent to part of 
maximally extended Schwarzschild solution in $\mathcal{S}$.}

It is also  interesting to note here that the covariant scale defined by equation \reff{EK} 
is equal to the Schwarzschild mass $m$.

%%%%%%%%%%%%%%%%%%%%%%%%%%%%%%%%%%%%%%%%%
\section{Spherically symmetric spacetimes with an almost vanishing Ricci Scalar}
%%%%%%%%%%%%%%%%%%%%%%%%%%%%%%%%%%%%%%%%%

From the previous section we know that for $f(R)$ gravity with $R = 0$, $f(0) = 0$ 
and $ \fp_{0} \neq 0$, all spherically symmetric vacuum spacetimes are locally 
isomorphic to a part of Schwarzschild solution. In \cite{amostBirkmatter}, the 
vacuum LRS II spacetime was perturbed by putting in a small amount of general 
matter that obeys weak and dominant energy conditions, to find out the amount of 
matter that can be introduced to the spacetime for the Jebsen-Birkhoff theorem to 
remain approximately true. Analogously, we investigate here the necessary conditions on the magnitude and spatial and temporal derivatives of the 
Ricci scalar, for the above theorem to remain approximately true. In this section 
we only deal with the static exterior background as it is astrophysically more 
interesting.

We have seen that the vacuum spherically symmetric spacetime with vanishing Ricci 
scalar has a covariant scale given by the Schwarzschild radius which sets up the 
scale for perturbation. Going by our description of the energy momentum tensor for 
vacuum LRS II spacetime in $f(R)$ gravity as consisting of curvature 
terms $\mu^R,\, p^R,\, \Pi^R$ and $Q^R $ and taking a static Schwarzschild background, then the set $\brac{R,\,\Theta,\,\Sigma}$ describe the 
first-order quantities (and are gauge-invariant according to the Stewart and Walker lemma \cite{Stewartlem}). 
Performing a series expansion of $f(R)$ in the neighbourhood of $R=0$ gives 
$f(R) = \fp_{0} \,R$ as a first-order term. Neglecting the higher order quantities in 
\reff{generaldensity}-\reff{generalstressTn} we get the following equations
\bea
\mu	 &=&  \frac{\fpp_{0}}{\fp_{0}} \bra{\hat{ X} + \phi \,X  } ~, 
			\label{muSS}\\
p &=& \frac{\fpp_{0}}{\fp_{0}} \bra{\ddot{R}  - \A\,  X 
			 - \frac23  \phi\, X   - \frac23 \hat{X} }  ~, 
			 \label{pSS} \\
Q &=& - \,\frac{\fpp_{0}}{\fp_{0}}  \bra{ \dot{X}-\A\, \dot{R}  }  ~,
		 \label{QSS}\\
\Pi &=& \frac{\fpp_{0}}{3\fp_{0}}   \bra{ 2 \hat{X} - \phi \, X  }   ~. 			\label{PiSS}
\eea
and
\be 
R \, \fp_0 =  3 \fpp_{0} \bra{ \hat{X} +( \A+ \phi)X - \ddot{R}} 
\label{traceSS}
\ee
for the trace. Thus we see that by perturbing the Ricci scalar in the neighbourhood 
of $R=0$ background, we are actually generating a `{\it curvature fluid}' having the 
above thermodynamic quantities. Therefore the situation here is similar to introducing 
small amount of matter on a Schwarzschild background in GR. In \cite{amostBirkmatter} 
the sufficient conditions for the {\it smallness} of these matter perturbations in order for 
the spacetime to remain almost Schwarzschild are given. These conditions in our case 
become
\begin{align}
&\bras{\frac{|R|}{K^{(3/2)}}, \,\frac{\fpp_{0}{}^{(1/2)} \,| \dot{R} |}{K^{(3/2)}},\,
\frac{ \fpp_{0} \,|\ddot{R}| }{K^{(3/2)}},\, \frac{\fpp_{0}{}^{(1/2)}\,|{X}|}{K^{(3/2)}},\right. \nn\\
& \left. \quad \qquad \frac{\fpp_{0}\,|{\hat{X}}|}{K^{(3/2)}},\,
\frac{\fpp_{0} \,|\dot{X}|}{K^{(3/2)}}} << C,
\label{cond1} 
\end{align}
and 
\be 
\bras{ \frac{\fpp_{0}{}^{3/2}\,  |\dddot{R}|}{K^{(3/2)}},\,
\frac{\fpp_{0}{}^{3/2} \, |\ddot{X} |}{K^{(3/2)}} }<< C \ .
\label{cond2} 
\ee 
Similarly to \cite{amostBirkmatter}, we also need to specify in what domain these equations will hold. This is important because eventually we will reach a radius $r$ 
where these inequalities may no longer hold. On the basis that in the real universe 
asymptotically flat regions are always of finite size, we will describe the local domain 
where our results will apply by \cite{amostBirk}, 
\begin{itemize}
\renewcommand{\labelitemi}{$-$}
 \item{Defining \emph{finite infinity}  ${\cal F}$ as a 2-sphere of radius $R_{\cal F} \gg C$ 
surrounding the star: this is infinity for all practical purposes \cite{EllisInf,StoegInf}.}
\item{Assuming the relations \reff{cond1}, \reff{cond2} hold in the domain $D_{\cal F}$ 
defined by $r_S < r < R_{\cal F}$ where $r_S > C$ is the radius of the surface of the star.}
\end{itemize}
We now linearise the field equations \reff{hatphi}-\reff{trace_SS} by neglecting the 
higher order quantities and we obtain the following equations for the 
first-order quantities
\bea
\hat\Sigma-\frac23\hat\Theta&\approx & -\,\frac32\phi\,\Sigma
+ \frac{\fpp_{0}}{\fp_{0}}  \bra{ \dot{X}-\A\, \dot{R}  }~,
\label{LE1} 
\eea
\bea
\dot\Theta&\approx&-\,\frac{\fpp_{0}}{2\fp_{0}} 
\bra{ 3 \ddot{R}-\hat{X} - (3\,\A+\phi)X}~,
\label{LE2} 
\eea
\bea
\dot\Sigma-\frac23\dot\Theta &\approx&  \frac{\fpp_{0}}{ \fp_{0}} 
\bras{\ddot{R} -X\bra{ \A + \frac12 \phi}}~,
\label{LE4} 
\eea
\bea
\dot\phi &\approx&\bra{\Sigma-\frac23\Theta}\bra{\A-\frac12\phi}
- \frac{\fpp_{0}}{\fp_{0}}  ( \dot{X}-\A \,\dot{R}  )~,
\label{LE3} 
\eea
\bea
\dot\E&\approx& \bra{\frac32\Sigma-\Theta}\E
+\phi \, \A\, \frac{\fpp_{0}}{2\fp_{0}}  \, \dot{R} 
 ~,
\label{LE5}
\eea
\bea
 \frac13 R \,\fp_{0}&\approx& \fpp_{0}\,\hat{X} - \fpp_{0}\,\ddot{R} 
 +   \bra{\phi+\A} \fpp_{0}\,X ~.
\label{trace5} 
\eea
From these equations we can see that if \reff{cond1} and \reff{cond2} are locally satisfied 
at any epoch, within the domain $D_{\cal F}$, then the spatial and temporal variation of 
the expansion $\Theta$ and the shear $\Sigma$ are of same order of smallness as the 
perturbations and derivatives of the Ricci scalar. In that case a timelike vector will not 
exactly solve the Killing equations \reff{psidot}-\reff{cons2} in general, although it may 
do so approximately. To see this explicitly, let us set $\Phi= 0$ in the Killing equation
\reff{Killing} 
\be 
\na_a(\Psi\,u_b) + \na_b(\Psi \,u_a) =0\;. 
\label{Killingb} 
\ee
and we once again try to solve the equation for a Killing vector of the form 
$\xi_a=\Psi \,u_a$ with an aim to see how close the $\xi_a $ is to 
Killing vector in the perturbed scenario. 

We consider the scalars constructed by multiplying 
the killing equation by the vectors $u^a$, $e^a$, the projection tensor $N^{ab}$ 
and utilise equation \reff{fullcov_e} and \reff{fullcov_u} to facilitate the calculation. 
We know that multiplying the Killing equation by $u^a\,u^b$ and $u^a\,e^b$ results in equations for which the solution of the scalar $\Psi$ always exists. The 
constraints obtained from multiplying the Killing equation by $e^a\,e^b$ and 
$N^{ab}$ only vanish if $\Theta=\Sigma=0$, however, we are considering here 
the perturbed case which is characterised by non-zero $\Theta$ and $\Sigma$. 
As a result not all the equations are completely solved in general. If we set up 
\reff{Killingb} as a symmetric tensor 
\be
K_{ab} := \nab_a(\Psi\,u_b) + \nab_b(\Psi \,u_a) \;. 
\label{Killing1}
\ee 
we can instead say that there always exists a non-trivial solution of the scalar $\Psi$ for 
which $|K_{ab}\,u^a\,u^b|$ and $|K_{ab}\,u^ae^b|$ vanishes and that 
$|K_{ab}\,e^a\,e^b|^2$ and $|K_{ab}\,N^{ab}|^2$ are non-zero since 
$\Theta$ and $\Sigma$ are non-zero. However, if the conditions
\begin{align}
\bras{\frac{|K_{ab}\,u^a\,u^b|^2}{K^{3/2}},\,\frac{|K_{ab}\,u^a\,e^b|^2}{K^{3/2}},
\frac{|K_{ab}\,e^a\,e^b|^2}{K^{3/2}},\,
\frac{|K_{ab}\,N^{ab}|^2}{K^{3/2}}}<<C 
\label{cond3} 
\end{align}
are satisfied, then we can say that $\xi_a = \Psi \,u_a$ is close to a
Killing vector and that the spacetime is approximately static.

Subtracting the background equation \reff{GaussCurv1} from \reff{GaussCurv}, 
we get 
\be 
\bra{\frac13 \Theta - \frac12 \Sigma}^{2}
~\approx  \frac{\fpp_{0}}{2 \fp_{0}} \, \phi X \ . 
\label{KE1} 
\ee
Similarly subtracting \reff{schwphi} from \reff{hatphi} we get 
\be
\bra{\frac13\Theta+\Sigma}\bra{\frac23\Theta-\Sigma}~\approx
 \frac{\fpp_{0}}{2 \fp_{0}} \bra{2\hat{X} +\phi\,X} \ . 
\label{KE2} 
\ee 
Using the above equations \reff{KE1} and \reff{KE2}, we immediately see that if 
\reff{cond1} is locally satisfied, then the following conditions are satisfied 
\be
|K_{ab}\,e^a\,e^b|^2=\Psi^2\bra{\frac13\Theta+\Sigma}^2\ll C\, K^{3/2}\ ,
\label{cons12} 
\ee 
\be
|K_{ab}\,N^{ab}|^2=\Psi^2\bra{\frac23\Theta-\Sigma}^2\ll C\, K^{3/2}\ .
\label{cons22} 
\ee 
It follows that there always exists a timelike vector that satisfies \reff{cond3}. 
This vector almost solves the Killing equations in the open 
set ${\mathcal S}$ in the domain $D_{\cal F}$ and hence the spacetime is 
{\it almost} static in $\mathcal{S}$. Moreover, the resultant field equations 
are the zeroth-order equations \reff{schwphi}-\reff{schwconstraint1} with an 
addition of $\mathcal{O}(\epsilon)$ terms.

%%%%%%%%%%%%%%%%%%%%%%%%%%%%%%%%%%%%%%%%%
\section{Almost spherically symmetric spacetime with vanishing Ricci scalar}%%%%%%%%%%%%%%%%%%%%%%%%%%%%%%%%%%%%%%%%%
In order to geometrically define an {\it almost spherically symmetric} spacetime, we 
begin by writing the {\it geodesic deviation equation} for a family of closely spaced 
geodesics on the 2-sheets with tangent vectors $\psi^a(v)$ and separation vectors 
$\eta^a(v)$ (where `$v$' is the parameter which labels the different geodesics) as 
\cite{amostBirkmatter},
\be
\psi^{e}\,\delta_{e}(\psi^{f}\, \delta_{f} \eta^a )=K(\psi^{a}\,\psi_{d}\, \eta^d 
-  \eta^a\,\psi_{c}\,\psi^{c} )\ .
\ee
We have used here the definition of the two dimensional Riemann curvature tensor 
equation \reff{2-RiemRic}.

We now define a vector $V^a$ by
\be
V^a =\psi^{e}\,\delta_{e}(\psi^{f}\, \delta_{f} \eta^a )
- K_0 (\psi^{a}\,\psi_{d}\, \eta^d -  \eta^a\,\psi_{c}\,\psi^{c} )~, 
\label{devVec}
\ee
where $K_0$ is the Gaussian curvature for a spherical sheet at any point $P$, which 
can be fixed by making the vector $V^a=0$ at that point. This vector vanishes for exact 
spherical 2-sheets in any open neighbourhood of $P$ but doesn't for non-spherical 
sheets. As a result, from the magnitude of $V^a (=\sqrt{ V_a\,V_a})$ we obtain a 
covariant measure of the deviation from the spherical symmetry.

We can now define an {\it almost spherically symmetric} spacetime in following
way \cite{varBirk}:

\textit{Any $C^3$ spacetime with positive Gaussian curvature everywhere, which 
admits a local 1+1+2 splitting at every point  is called an \emph{almost 
spherically symmetric} spacetime, if and only if the following quantities are 
either zero or much smaller than the scale defined by the modulus of the 
proportionality constant in equation \reff{EK}:
\begin{itemize}
\renewcommand{\labelitemi}{$-$}
\item{The magnitude of all the 2-vectors (defined by $\sqrt{\psi_a\psi^a}$)
and PSTF 2-tensors (defined by $\sqrt{\psi_{ab}\psi^{ab}}$).}
\item {The magnitude of vector $V^a$ defined above in \reff{devVec}.}
\end{itemize}}

We have seen that subject to the conditions \reff{cond1} 
and \reff{cond2}, on any spherically symmetric local domain $D_{\cal F}$, the spacetime 
remains ``almost'' Schwarzschild for all the $f(R)$-theories that admit a Schwarzschild 
background, (that is, a background characterised by a vanishing Ricci scalar with $f(0) = 0$ 
and $ \fp_{0} \neq 0$). We now wish to see to what extent the conditions hold when 
we perturb this geometry. 

As previously stated, the sheet will be a genuine two surface if and only if the commutator 
of the time and hat derivative do not depend on any sheet component and the sheet 
derivatives commute in \reff{comm1} and \reff{comm4}. 
Following from the definition of almost spherical symmetry, in the perturbed scenario we 
will require the sheet to be an almost genuine 2-surface such that the commutator of the 
time and hat derivative {\it almost} do not depend on any sheet component and the sheet 
derivatives {\it almost} commute. In that case we see from \reff{comm1} and \reff{comm4} 
that the scalars $\Omega$ and $\xi$ will be of the same order of smallness as the 
other vectors and PSTF 2-tensors on the sheet. Furthermore, from the constraint equation 
\begin{align}
\delta_a\Omega^a+\veps_{ab}\delta^a\Sigma^b=\bra{2\A-\phi}\Omega
    -3\xi\,\Sigma+\veps_{ab}\zeta^{ac}\Sigma^b_{~c} +\H~,
    \label{divOmeganl1}
\end{align}
we see that the scalar $\H$ is of the same order of smallness. Dealing 
once again with the static exterior background, we now have it that the set of 1+1+2 variables
\bea
&\bras{R,\,\Theta, \,\Sigma, \,\Omega, \,\H, \,\xi, \,\A^{a},
\, \Omega^{a}, \,\Sigma^{a}, \alpha^a,\,\right. \nn\\
& \left.  \qquad \hatn^a,\, \E^{a}, \,\H^{a},\,\Sigma_{ab }, 
\,\E_{ab},\,\H_{ab},\,\zeta_{ab}}\,,
\label{firstorder}
\eea
are all of ${\mathcal O}(\epsilon)$ with respect to the invariant scale. We shall treat these 
variables along with their derivatives and the dot - $`\dot{\phantom{x}}$' and delta - $`\delta$' 
derivatives of $\brac{\A,\, \E,\,\phi}$ as first-order relative to the background terms. 

Performing a series expansion of $f(R)$ in the neighbourhood of $ R=0 $ and linearising by 
neglecting all products of first order quantities in \reff{generaldensity}-\reff{generalstressTn}, 
we obtain
\bea
\mu
&\approx&\frac{\fpp_{0}}{\fp_{0}}  \bra{\hat{ X} + \phi \, X  + \delta^{2}R  } ~, 
 \label{almostdensity}\\
p
&\approx& \frac{\fpp_{0}}{\fp_{0}}\bras{\ddot{R}  - \A \, X -\frac23 \bra{ \phi \, X 
+  \hat{X} + \delta^{2} R}} ~,
 \label{almostpressure}\\
Q
&\approx&- \,\frac{\fpp_{0}}{\fp_{0}} \bra{ \dot{X}-\A \,\dot{R} } ~,
 \label{almostfluxSc}\\
 Q_{a}
&\approx&-\,  \frac{\fpp_{0}}{\fp_{0}} \, \delta_{a} \dot{R} ~,  
 \label{almostfluxVc}\\
 \Pi  
&\approx& \frac{\fpp_{0}}{3\fp_{0}}   \bra{ 2 \hat{X} 
 - \phi \, X - \delta^{2} R } ~,
 \label{almoststressSc}\\
 \Pi_{a} &\approx& \frac{\fpp_{0}}{\fp_{0}} \bra{  \delta_{a} X 
- \frac12 \phi \, \delta_{a} R } ~,
 \label{almoststressVc}\\
 \Pi_{ab}
&\approx&\frac{\fpp_{0}}{\fp_{0}} \,\delta_{\lb a} \delta_{b \rb} R~.
 \label{almoststressTn}
\eea
Linearising the field equations (48-81) in \cite{chris} and substituting in equations 
\reff{almostdensity} - \reff{almoststressTn} we obtain:\\
\\
{\it Evolution equations}\\
\\
The evolution equations for $\xi$ and $\zeta_{\lb ab\rb}$
are:
 \be
\dot{\xi} = \bra{\A-\frac12\phi}\Omega
+\frac12\veps_{ab}\delta^a\alpha^b +\frac12 \H ~;
\label{dotxinl}
\ee
\be
\dot\zeta_{\lb ab\rb}=\bra{\A-\frac12\phi}\Sigma_{ab}
+\delta_{\lb a}\alpha_{b\rb} 
 -\veps_{c\lb a}\H_{b\rb}^{~~c}~;
 \label{dotzetanl}
\ee
Vorticity evolution equation:
\be
\dot\Omega = \frac12\veps_{ab}\delta^a\A^b+\A\,\xi~,
\ee
\be
\dot\Omega_{\bar a} + \frac12\veps_{ab}\hat\A^b =
  \frac12\veps_{ab} \bra{ \delta^b \A - \A \,\hatn^b - \frac12 \phi\, \A^b}~;
\ee
Shear evolution:
\bea
\dot\Sigma_{\bar a}-\frac12 \hat{\A}_a 
&=& \frac12 \delta_a\A+\bra{\A-\frac14\phi}\A_a
    + \frac12 \A \,\hatn_a -\E_a  \nn\\
&&  + \frac{\fpp_{0}}{2\fp_{0}} \bra{  \delta_{a} X 
    - \frac12 \phi\, \delta_{a} R  }~,
\eea
\be
\dot\Sigma_{\lb ab\rb} = \delta_{\lb a}\A_{b\rb} 
    +\A\,\zeta_{ab}-\E_{ab}
    +\frac{\fpp_{0}}{2\fp_{0}}  \,\delta_{\lb a} \delta_{b \rb} R~;
\ee
Magnetic Weyl evolution:
\be
\dot\H = -\,\veps_{ab}\delta^a\E^b-3\xi\,\E~,
\ee
\bea
\dot\H_{\bar a} &=&
   -\,\frac32 \E\, \veps_{ab} \A^b -\frac12\veps_{ab} \delta^b\E 
 -\frac12 \bra{\phi - 2\A} \veps_{ab} \E^b \nn\\
&&+ \veps_{c\lb d} \delta^d \E_{a\rb}^{~~c} 
 - \E \, \frac{\fpp_{0}}{4 \fp_{0}} \, \veps_{ab}\delta^b R~,
\eea
\bea
\dot\H_{\lb ab\rb}+\veps_{c\lb a}\hat\E_{b\rb}^{~~c} &=&
	\veps_{c\lb a}\delta^c\E_{b\rb}
	+\frac32 \E\,\veps_{c\lb a}\zeta_{b\rb}^{~~c} \nn\\
	&&-\bra{\frac12\phi+2\A}\veps_{c\lb a}\E_{b\rb}^{~~c}~;
\eea
Electric Weyl evolution:
\begin{align}
\dot\E_{\bar a} + \frac12 \veps_{ab} \hat\H^b  =&
	 \frac34 \E  \bra{\veps_{ab}\Omega^b + \Sigma_a-2 \alpha_a}+\frac34\veps_{ab} \delta^b\H \nn\\
	 & - \bra{\frac14 \phi+ \A} \veps_{ab}\H^b
	  +\frac12\veps_{bc} \delta^b\H^c{}_a~,
\end{align}
\bea
\dot\E_{\lb ab\rb}-\veps_{c\lb a}\hat\H_{b\rb}^{~~c}&=&
	-\,\veps_{c\lb a}\delta^c\H_{b\rb}
	 -\frac32 \E\, \Sigma_{ab}\nn\\
	 &&+\bra{\frac12\phi+2\A}\veps_{c\lb a}\H_{b\rb}^{~~c}~;
\eea
Evolution equation for $\hat{e}_a$:
\bea
\dot a_{\bar a}-\hat\alpha_{\bar a}&=&
    \bra{\frac12\phi+\A}\alpha_a
    -\bra{\frac12\phi-\A}\bra{\Sigma_a+\veps_{ab}\Omega^b}\nn\\
	 &&+\,\veps_{ab}\H^b +\frac{\fpp_{0}}{2 \fp_{0}}\,  \delta_{a} \dot{R}~.
  \label{hatalphanl}
\eea
\\
{\it Propagation equations}
\be
\hat\xi=-\,\phi\,\xi+\frac12\veps_{ab}\delta^a \hatn^b 
\label{hatxinl}~;
\ee
\be
\hat\zeta_{\lb ab\rb}=-\,\phi\,\zeta_{ab}
	+\delta_{\lb a} \hatn_{b\rb } -{\cal E}_{ab} 
	- \frac{\fpp_{0}}{2 \fp_{0}}  \,\delta_{\lb a} \delta_{b \rb} R~;
	\label{hatzetanl}
\ee
Shear divergence:
\bea
\hat\Sigma_{\bar a}-\veps_{ab}\hat\Omega^b&=&\frac12\delta_a\Sigma
    +\frac23\delta_a\theta - \veps_{ab}\delta^b\Omega-\frac32\phi\,\Sigma_a
    -\delta^b\Sigma_{ab} \nn\\
    &&+\,\bra{\frac12\phi+2\A}\veps_{ab}\Omega^b
    +\frac{\fpp_{0}}{\fp_{0}}\,\delta_{a} \dot{R}~,
\eea
\be
\hat\Sigma_{\lb ab\rb}=\delta_{\lb a}\Sigma_{b\rb} 
-\veps_{c\lb a}\delta^c\Omega_{b\rb}
    -\frac12\phi\,\Sigma_{ab} -\veps_{c\lb a}\H_{b\rb}^{~~c}~;
\ee
Vorticity divergence equation:
\bea
\hat\Omega=-\,\delta_a\Omega^a+\bra{\A-\phi}\Omega ~;
\label{hatOmSnl}
\eea
Electric Weyl Divergence:
\be
\hat\E_{\bar a} = \frac12\delta_a\E
    -\delta^b\E_{ab} -\frac32 \E\, a_a
    -\frac32\phi \,\E_a +\E \, \frac{\fpp_{0}}{4 \fp_{0}}\,\delta_a R~;
\ee
Magnetic Weyl divergence:
\be
\hat\H= -\,\delta_a\H^a -\frac32\phi\,\H -3\E\, \Omega~,
\ee
 \be
\hat\H_{\bar a} =
	\frac12\delta_a\H-\delta^b\H_{ab}
  	  +\frac32 \E \bra{\Omega_a-\veps_{ab}\Sigma^b }
  	  -\frac32\phi\,\H_a  ~.
	   \label{hatHnl}
\ee
Finally we have the linearised curvature trace equation 
\be
 \frac13 R = \frac{\fpp_{0}}{ \fp_{0}}\bras{\hat{X} - \ddot{R} 
 +   \bra{\phi+\A} X+\delta^{2} R}~.
\label{Curvetracenl} 
\ee
From the evolution equations \reff{dotxinl} - \reff{hatalphanl},  it is evident that if the 
background is static with $\Sigma = \Theta = 0$ or ``almost static'' with 
$\Sigma = \Theta = {\mathcal O}(\epsilon)$, the time derivatives of the first-order quantities 
at a given point are all of the same order of smallness as the variables themselves. Hence if at a 
given epoch these quantities are of $ {\mathcal O}(\epsilon)$, then there exists an open 
set ${\mathcal S}$ in the domain  $D_{\cal F}$ where these quantities continue to be of the 
same order. 

This time if we project the {\it Killing equation} \reff{Killing} for a Killing vector of the form 
$\xi_a=\Psi \,u_a$, with $N^a_c\,u^b$, 
$N^a_c\,e^b$ and $N^a_c\,N^b_d$, we obtain the following additional constraints 
on the 2-sheet: 
\bea
-\delta_c\Psi+\Psi\,\A_c&=&0\;,
\label{delpsi}\\
\Psi\,\Sigma_c&=&0\;,
\label{delphi}\\
\Psi\,\Sigma_{cd}&=&0\;.
\label{cons31} 
\eea 
The solution of \reff{delpsi} always exists and as we have just seen, the LHS of equations 
\reff{delphi} and \reff{cons31} remains ${\mathcal O}(\epsilon)$ in  ${\mathcal S}$.
Hence a timelike vector almost solves the  Killing equations, making the spacetime 
almost static. 
%%%%%%%%%%%%%%%%%%%%%%%%%%%%%%%%%%%%%%%%%%%%%%%%%%%%%%%%%%%%%%%%%%%%
\section{Local stability of Jebsen-Birkhoff like theorem}
%%%%%%%%%%%%%%%%%%%%%%%%%%%%%%%%%%%%%%%%%%%%%%%%%%%%%%%%%%%%%%%%%%%%

Let us now combine the results obtained in the previous two sections. Consider any $f(R)$ theory of gravity which 
admits a Schwarzschild background and consider the following sets of scalars:
\begin{align}
&\bras{\frac{|R|}{K^{(3/2)}}, \,\frac{\fpp_{0}{}^{(1/2)} \,| \dot{R} |}{K^{(3/2)}},\,
\frac{ \fpp_{0} \,|\ddot{R}| }{K^{(3/2)}},\, \frac{\fpp_{0}{}^{(1/2)}\,|{X}|}{K^{(3/2)}},\right. \nn\\
& \left. \quad \qquad \frac{\fpp_{0}\,|{\hat{X}}|}{K^{(3/2)}},\,
\frac{\fpp_{0} \,|\dot{X}|}{K^{(3/2)}},\frac{\fpp_{0}{}^{3/2}\,  |\dddot{R}|}{K^{(3/2)}},\frac{\fpp_{0}{}^{3/2} \, |\ddot{X} |}{K^{(3/2)}}} .
\label{scalars} 
\end{align}
If these scalars locally satisfy \reff{cond1}, \reff{cond2} and their sheet derivatives are of same order of smallness as themselves 
at any epoch within the domain $D_{\cal F}$, then there exists an open set ${\mathcal S}$ in $D_{\cal F}$ where 
the conditions continue to hold. Consequently then there will exist a timelike/spacelike vector that almost solves the Killing equations 
in ${\mathcal S}$ and hence the spacetime will be "almost" Schwarzschild.  Hence we have demonstrated an important result : {\em For $f(R)$ gravity, where the function $f$ is of class $C^3$ at $R=0$,with $f(0) = 0$ 
and $\fp(0)\ne 0$, any almost spherically symmetric solution with almost vanishing Ricci scalar in 
empty space in an open set $\mathcal{S}$, is locally almost equivalent to part of 
maximally extended Schwarzschild solution in $\mathcal{S}$.}

We would like to emphasise here that the size of the open set ${\mathcal S}$ depends on the 
parameters of theory (namely the quantity $\fpp(0)$) and the covariant scale (which is the 
Schwarzschild mass of the star) and we can always tune the parameters of the theory such that 
the perturbations continue to remain small for a time period which is greater than the age of the 
universe. The above result shows that the local spacetime around almost spherical stars will be stable in the 
regime of linear perturbations in these modified gravity theories.
%%%%%%%%%%%%%%%%%%%%%%%%%%%%%%%%%%%%%%%%%%%%%%%%%%%%%%%%%%
\section{Discussion}
%%%%%%%%%%%%%%%%%%%%%%%%%%%%%%%%%%%%%%%%%%%%%%%%%%%%%%%%%%%

In this paper we used the 1+1+2 covariant perturbation formalism to prove a Jebsen-Birkhoff like theorem for $f(R)$ theories of gravity in order to determine the conditions required for the existence of the Schwarzschild solution in these theories. We then discussed under what circumstances we can covariantly set up the fundamental scale in the problem and perturbed the vacuum spacetime with respect to this scale to find the stability of the theorem.

What emerges from this analysis is the important result that there exists a non-zero 
measure in the parameter space of $f(R)$ theories of gravity for which the Jebsen-Birkhoff like theorem remains stable under generic perturbations. This result applies locally and therefore does not depend on specific boundary conditions used for solving the perturbation equations. A detailed analysis of generic linear perurbations of the Schwarzshild solution will be presented elsewhere, which supports the work presented in this paper.

%%%%%%%%%%%%%%%%%%%%%%%%%%%%%%%%%%%%%%%%%%%%%%%%%%%%%%%%%%%%%


\begin{thebibliography}{99}
%%%%%%%%%%%%%%%%%%%%%%%%%%%%%%%%%%%%%%%%%%%%%%%%%%%%%%%%%%%%%
\bibitem{birkI}
J.T. Jebsen, ``\"{U}ber die allgemeinen kugelsymmetrischen L\"{o}sungen der Einsteinschen Gravitationsgleichungen im Vakuum'', {\it Arkiv for Matematik, Astronomi och Fysik}, {\bf 15}, 18 (1921). ~Reprinted as a Golden Oldie: {\it Gen. Relativ. Gravit.} {\bf 37}, no. 12, 2253 (2005).

\bibitem{birkII}
G. D. Birkhoff,  ``Relativity and Modern Physics'', (Cambridge, MA: Harvard University Press: 1923).

\bibitem{birkIII}
S. W. Hawking and G. F. R. Ellis, ``The Large Scale Structure of Spacetime'', (Cambridge University Press: 1973).

\bibitem{CDGN}
  T.~Clifton, P.~Dunsby, R.~Goswami and A.~M.~Nzioki,
  Phys.\ Rev.\ D {\bf 87}, no. 6, 063517 (2013)
  [arXiv:1210.0730].

\bibitem{scalar-tensor}
 Valerio Faraoni, ``Cosmology in Scalar-Tensor Gravity'', (Fundamental Theories of Physics, Springer: 2004).

\bibitem{FS} T. P. Sotiriou and V. Faraoni, ``Black holes in scalar-tensor gravity'', {\it Phys. Rev. Lett.} {\bf 108}, 081103 (2012) [arxiv:1109.6324].\bibitem{amostBirk}
R. Goswami and  G. F. R. Ellis, ``Almost Birkhoff Theorem in General Relativity'', {\it Gen. Relativ. Gravit.} \textbf{43}, 2157 (2011)
[arXiv:1101.4520]. 

\bibitem{amostBirkmatter}
R. Goswami and G. F. R. Ellis, ``Birkhoff theorem and matter'', {\it Gen. Relativ. Gravit.} \textbf{44}, 2037 (2012) [arXiv:1202.0240v1].

\bibitem{varBirk}
G. F. R. Ellis, and R. Goswami,  ``Variations on Birkhoff's theorem'',  {\it Gen. Relativ. Gravit.} \textbf{45}, 2123 (2013) [arXiv:1304.3253v1].

\bibitem{extension}
C. A. Clarkson and R. K. Barrett, ``Covariant perturbations of Schwarzschild black holes'', {\it Class.\ Quant.\ Grav.} {\bf 20}, 3855 (2003) [arXiv:gr-qc/0209051v3].

\bibitem{Gary}
G. Betschart and C. A. Clarkson, ``Scalar field and electromagnetic perturbations on Locally Rotationally Symmetric spacetimes'', {\it Class.\ Quant.\ Grav.} {\bf 21} 5587 (2005) [arXiv:gr-qc/0404116v3].

\bibitem{chris}
C. Clarkson, ``A covariant approach for perturbations of rotationally symmetric spacetimes'', {\it Phys.\ Rev.\ D.} {\bf 76}, 104034 (2007) [arXiv:0708.1398v1].

\bibitem{BCMD} 
  C.~A.~Clarkson, M.~Marklund, G.~Betschart and P.~K.~S.~Dunsby,
  Astrophys.\ J.\  {\bf 613}, 492 (2004)
  [astro-ph/0310323].

\bibitem{StewartLRS}
J.M. Stewart and G. F. R. Ellis, ``Solutions of Einstein's Equations for a Fluid Which Exhibit Local Rotational Symmetry'', {\it 
J. Math. Phys.} {\bf 9}, 1072 (1968).

\bibitem{MarklundLRS}
M. {Marklund} and M. Bradley, ``Invariant construction of solutions to Einstein's field equations - LRS perfect fluids II'', ', {\it Class.\ Quant.\ Grav.} {\bf 16}, 1577 (1999) [arXiv:gr-qc/9808062].

\bibitem{EllisLRS} 
G. F. R. Ellis, ``The dynamics of pressure-free matter in general relativity", {\it J. Math. Phys.} {\bf 8}, 1171 (1967). ~H. van Elst  and G. F. R. Ellis, ``The Covariant Approach to LRS Perfect Fluid Spacetime Geometries'', {\it Class.\ Quant.\ Grav.} {\bf 13}, 1099 (1996) [arXiv:gr-qc/9510044v1].

\bibitem{Stewartlem} 
J.M. Stewart and M. Walker, ``{Perturbations of spacetimes in general relativity}'', {\it Proc.Roy.Soc.Lond.} {\bf A341}, 29 {1974}.

\bibitem{EllisInf} 
G. F. R. Ellis, ``Relativistic Cosmology: Its Nature, Aims and Problems'', {\it Gen. Relativ. Gravit.} \textbf{9}, 215 (1984).

\bibitem{StoegInf} 
G. F. R. {Ellis} and W. R. {Stoeger}, ``{The evolution of our local cosmic domain: effective causal limits}'', {\it MNRAS} {\bf 398}, 1527 (2009) [arXiv:1001.4572v1].

\bibitem{Covariant}
G. F. R. Ellis \& M. Bruni, ``Covariant and gauge invariant approach to density fluctuations'', {\it Phys Rev D} {\bf 40} 1804 (1989);  M. Bruni,  P. K. S. Dunsby \& G. F. R. Ellis, ``Cosmological perturbations and the physical meaning of gauge-invariant variables'', {\it Ap. J.}, {\bf 395} 34 (1992).

\end{thebibliography}
\end{document}